\documentclass[prl,floatfix,twocolumn,showpacs,amsmath,amssymb]{revtex4}
\usepackage{graphicx,color}
\usepackage{dcolumn}
\usepackage{bm}
\begin{document}
\title{d-wave pairing in lightly doped Mott insulators}
\author{Evgeny Plekhanov, Federico Becca, and Sandro Sorella}
\affiliation{
INFM-Democritos, National Simulation Centre, and SISSA I-34014 Trieste, Italy
}
\date{\today}
\begin{abstract}
We define a suitable quantity $Z_c$ that measures the pairing strength 
of two electrons added to the ground state wave function by means of 
the anomalous part of the one-particle Green's function.
$Z_c$ discriminates between systems described by one-electron states, 
like ordinary metals and band insulators, for which $Z_c=0$, and systems 
where the single particle picture does not hold, like superconductors and 
resonating valence bond insulators, for which $Z_c \ne 0$. 
By using a numerically exact projection technique for the Hubbard model at
$U/t=4$, a finite value of $Z_c$, with d-wave symmetrry, is found at 
half filling and in the lightly doped regime, thus 
emphasizing a qualitatively new feature coming from electronic correlation.
\end{abstract}
\maketitle

Since the discovery of high-temperature (HTc) superconductors,
the question whether a strongly correlated system, containing 
{\it only} repulsive interactions, may display a superconducting ground 
state, has been intensively debated until now, mainly  because  
HTc superconductors are certainly strongly correlated and 
their critical temperature cannot be explained by a standard electron-phonon 
mechanism. Indeed, any mean-field approximation fails to explain
superconductivity without an electron-electron attraction, that could be
either explicit, like in the attractive Hubbard model, or mediated by some 
boson, like in the BCS theory.
On the other hand, it is well known that correlated wave functions,
that highly improve the mean-field energies, usually favor
superconductivity.~\cite{nakanisi,gros,randeria,ivanov,sorella}
Recently, the approach based on projected superconducting wave functions has
been recently renewed, as it is possible to reproduce many important 
experimental aspects just by {\it assuming} that the ground state is described
by a projected BCS wave function.~\cite{randeria,vanilla}
Within this scheme, the superconductivity is ``hidden'' in the insulating
state, where phase coherence is inhibited by the charge gap, and it is 
indeed stabilized by a small amount of doping.
Unfortunately, the  validity of this scenario for the actual
ground state of a microscopic model remains an highly debated and 
controversial issue, despite a huge amount of numerical 
work.~\cite{sorella,tklee,gubernatis,zhang,white}

In this letter, we present a numerical study in favor of the mentioned 
scenario. We consider the single-band Hubbard model on $L$-site clusters:
\begin{equation} \label{model}
H= -t \sum_{\langle i,j \rangle, \sigma} c^\dag_{i,\sigma} c_{\sigma,j} + h.c.
+U \sum_i  n_{i,\uparrow} n_{i,\downarrow} -\mu N,
\end{equation}
where $c^{\dag}_{i,\sigma}$ creates an electron with spin $\sigma$ at the
site $i$, $n_{i,\sigma}= c^{\dag}_{i,\sigma} c_{i,\sigma}$ is the density
operator at the site $i$, $\mu$ is the chemical potential, and
$N=\sum_{i, \sigma} n_{i,\sigma}$ is the total number of particles.
We use square lattices tilted by $45^\circ$ with $L^2=2 l^2$ and $l$ odd
with periodic boundary conditions,
so that the half-filled case ($N=L$) has a non-degenerate ground state
even at $U=0$. This condition strongly reduces the size effects,
particularly important in two dimensions (2D).

Our purpose is to study the anomalous part of the equal-time Green's function
at zero temperature:
\begin{equation}
F_k^{BSP}=\langle \Psi_0| c^{\dag}_{k,\uparrow} c^{\dag}_{-k,\downarrow}
|\Psi_0 \rangle,
\end{equation}
where $|\Psi_0 \rangle$ is the ground state of the Hamiltonian $H$.
Obviously, $F_k^{BSP}$ can be non zero only in the thermodynamic limit, where,
because of the broken symmetry phenomenon (BSP), $|\Psi_0 \rangle$ may not
have a definite number of particles.
On a finite system BSP does not occur and $F_k^{BSP}$ is always zero, 
the ground state $|\Psi_0 \rangle$ having always a definite number of 
particles $N$.
Therefore, on finite systems, it is useful to consider a related quantity:
\begin{equation} \label{final}
F_k = \langle \Psi_0^{N+2}| c^{\dag}_{k,\uparrow} c^{\dag}_{-k,\downarrow}
| \Psi_0^N \rangle,
\end{equation}
where $|\Psi_0^N \rangle$ is the ground state with $N$ particles.
Following well known concepts of BSP, by adding to the Hamiltonian a small 
anomalous term, that breaks the symmetry, proportional to $\delta$, and 
taking the limit $\delta \to 0$, after the thermodynamic limit, it is easy 
to convince  that the finite-size expression $F_k$ converges to $F_k^{BSP}$.

In order to evaluate $F_k$, we consider a trial BCS
wave function $|\Psi_G \rangle$ containing only components with even
number of particles:
\begin{equation}\label{wf}
|\Psi_G  \rangle = {\cal J} {\rm exp}
\left( \sum_{k} f_{k} c^{\dag}_{k,\uparrow} c^{\dag}_{-k,\downarrow} \right)
|0\rangle,
\end{equation}
where $|0 \rangle$ is the vacuum,
$f_k=\Delta_k/(\epsilon_k-\mu_0+E_k)$, $\epsilon_k=-2t(\cos k_x+\cos k_y)$
is the free-electron dispersion, $\mu_0$ is a variational parameter playing
the role of the chemical potential of the BCS Hamiltonian,
$\Delta_k$ is the corresponding gap function chosen to have either a d-wave
symmetry $\Delta_k=\Delta(\cos k_x-\cos k_y)$ or an s-wave symmetry
$\Delta_k=\Delta$, and $E_k=\sqrt{(\epsilon_k-\mu_0)^2+\Delta_k^2}$. 
The correlation factor 
${\cal J}=e^{-g \sum_i n_{i,\uparrow} n_{i,\downarrow}}$, $g$ being a 
variational parameter, partially projects out expensive energy configurations
with doubly occupied sites.
In the following, we assume that, as verified on small clusters, for an
appropriate gap function symmetry, the $N$-particle ground state of $H$
have non-vanishing overlap with $|\Psi_G \rangle$, at least for $N=N^*$ 
and $N=N^*+2$, where $N=N^*$ represents a closed shell density.
Then, on each finite system, we compute for fixed imaginary time $\tau$:
\begin{equation} \label{fktau}
F_k(\tau)= \frac{\langle \Psi_G| e^{-\tau H/2}
c^{\dag}_{k,\uparrow} c^{\dag}_{-k,\downarrow} e^{-\tau H/2} |\Psi_G \rangle}
{\langle \Psi_G| e^{-\tau H} |\Psi_G \rangle}.
\end{equation}
$F_k(\tau)$ will be finite even on a finite size,
because the trial function $|\Psi_G \rangle$ contains sectors with different
$N$, namely $|\Psi_G \rangle= \sum_N a_N |\Psi_G^N \rangle$. 
Then for large $\tau$:
\begin{equation} \label{fk}
F_k(\tau)= F_k \frac{b_{N^*+2}}{b_{N^*}} e^{-\tau \Delta_c/2}
\end{equation}
where $N^*$ is chosen by fixing the true chemical potential $\mu$,
$\Delta_c$ is the energy gap between the states with $N^*$ and
$N^*+2$, and $b_N= \langle \Psi_G^N| \Psi_0^N \rangle a_N$.
Therefore, in order to estimate  $F_k$, the overall factor
$\gamma=b_{N^*+2}/b_{N^*} e^{-\tau \Delta_c/2}$ of Eq.~(\ref{fk}) must
be computed by evaluating the average number of particles
$\langle N \rangle= N^*+2 \gamma^2/(1+\gamma^2)$.
An efficient choice of $|\Psi_G \rangle$ is obtained by tuning $\mu_0$ 
very close to the energy level ${\bar \epsilon_k}$ with
$N^*+2$ particles, and $\Delta \ll |{\bar \epsilon_k} -\mu_0|$.
In this limit, $|\Psi_G \rangle$ essentially contains only two components in 
the sectors with $N=N^*$ and $N=N^*+2$ and the imaginary time projection can be
done without paying much attention to the chemical potential, the
number of particles being conserved by $H$.

The superconducting order parameter $P$ is simply related to the
short-distance component of the real-space Fourier transform
$F_R= 1/L \sum_k e^{i k R} F_k$, namely
$P_d =2  F_\eta$, with $\eta=(\pm 1,0)$ or $\eta=(0,\pm 1)$ for a d-wave
superconductor with $\Delta_k=\Delta(\cos k_x - \cos k_y)$,
and $P_s = F_\eta$ 
with $\eta=(0,0)$ for an s-wave superconductor with $\Delta_k=\Delta$.
Based on the resonating-valence bond (RVB) theory,~\cite{rvb} we want to
define a quantity, closely related to $F_k$, that makes sense
also when $P_d=0$, as in the insulating case.
Indeed, according to the RVB theory, the insulator already contains some 
sort of pairing, the electrons being paired in RVB singlets.
Following this paradigm, we normalize the anomalous pairing function
in real space and define the quantity:
\begin{equation} \label{defgr}
g_R= \frac{F_R}{\sqrt{\sum_{R^\prime} F_{R^\prime}^2}},
\end{equation}
where the sum is for all distances in the lattice (compatible with the
symmetry of $F_R$).
Notice that, also for an infinite system with a charge gap, like the
half-filled Hubbard model, though the anomalous average $F_R$ is zero,
a finite ratio $g_R$ is still possible.
$g_R^2$ may be interpreted as the probability function of two electrons added
into a singlet state at a distance $R$ and, therefore, determines the
pair function, which is the Cooper pair if $P>0$.
For a BCS superconductor with a full gap (e.g., s-wave) $g_R$
is localized with a well defined coherence length, whereas if the pairing
function has nodes, like in the d-wave case, $g_R$ decays with a power law,
but, nonetheless, it exists a {\it finite} amplitude of the pairing function
at short distance.
In strong-coupling superconductivity, the coherence length is expected to be
small, and the most important contribution to the pairing function is at
the shortest distance $\eta$ allowed by the symmetry of the pairing function.
Therefore, we define a fundamental quantity, the {\it pairing strength},
that we denote by:
\begin{equation}
Z_c= | g_\eta |.
\end{equation}

\begin{figure}
\includegraphics[width=0.50\textwidth]{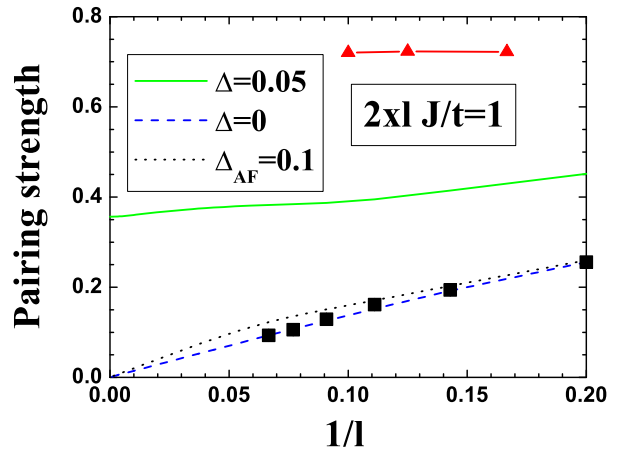}
\vspace{-1.2cm}
\caption{\label{plotbcs}
Size scaling of the d-wave pairing strength $Z_c$ for different wave functions
at half-filling: the uncorrelated BCS wave function with $\Delta=0.05$, 
the free-electron wave function ($\Delta \to 0$), the wave function with a 
finite antiferromagnetic order parameter [obtained by adding to the 
$\Delta \to 0$ BCS Hamiltonian a further antiferromagnetic mean field term 
$(-1)^i \Delta_{AF} (n_{i,\uparrow}-n_{i,\downarrow})$], and
the Gutzwiller wave function with $g=1$ (full squares).
The same quantity for the exact ground state of the two-chain $t{-}J$ ladder 
at low doping is also shown (full triangles).
}
\end{figure}

As shown in Fig.~\ref{plotbcs}, in the thermodynamic limit, $Z_c$ is finite
in an uncorrelated d-wave superconductor and, instead, it vanishes for free 
electrons, for a weakly-correlated Fermi liquid, described by a Gutzwiller 
wave function, and for a spin-density wave state, with a finite 
antiferromagnetic order parameter.
By introducing the pairing strength $Z_c$, we provide a general and
quantitative definition of pairing, which holds not only for simple
superconductor systems, but also for non-BCS systems.
A simple example is the two-leg ladder (see Fig.~\ref{plotbcs}),
where off-diagonal long-range order is suppressed by one-dimensional 
quantum fluctuations, but, nevertheless, $Z_c$ remains finite, showing that 
pairing is well defined also in this system, as widely 
accepted.~\cite{poilblanc,scalapino,dagotto}

\begin{figure}
\includegraphics[width=0.50\textwidth]{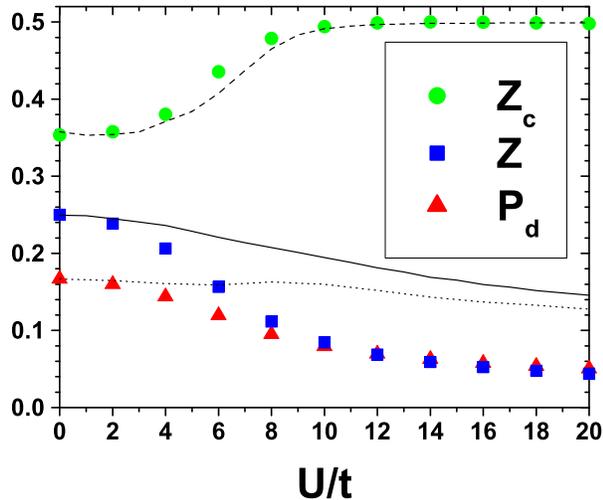}
\vspace{-1.2cm}
\caption{\label{small}
The d-wave pairing strength $Z_c$ (circles) as a function of $U$ in an 
$L=18$ site cluster with  $N^*=L$. The quasiparticle weight $Z_{N}$ for $N=16$ 
(squares) at the momentum
$(2 \pi/3,0)$ and the short-distance anomalous average $P_d$ (triangles) are
also shown. Continuous ($Z_N$), long dashed ($Z_c$) and dotted ($P_d$) 
lines correspond to the variational calculation with 
the variational wave function of Eq.~(\ref{wf}) with an optimized $\Delta_k$ 
and projected onto the subspaces with $16$ and $18$ electrons.
}
\end{figure}

There is also another important reason to study the pairing strength
$Z_c$ instead of the order parameter $P$: in a strongly correlated system 
the value of the quasiparticle weight $Z_N$, defined
as $Z_N=|\langle \Psi_0^{N+1}| c^{\dag}_k |\Psi_0^N \rangle|^2$,
can be very small~\cite{schrieffer,martins}
and $P$, if finite, is expected to be at most of the same order. 
Therefore, whenever $Z_N$ is very small, it is very difficult to detect a 
non-zero value of the anomalous average $P$.
The suppression of d-wave pairing obtained in previous
calculations~\cite{gubernatis,zhang} can be explained by the fact that the
quasiparticle weight decreases very rapidly with $U$ (see below and 
Ref.~\cite{ogata}).
The question of a finite anomalous average of order $Z_N$ could be still 
compatible with the published numerical calculations and it is beyond the 
scope of this work.
For this reason it is very difficult, at the present time, to detect
pairing by studying directly the order parameter or the pairing correlations.
On the other hand, the pairing strength $Z_c$, being a ratio of two quantities
of the same order, is not affected by the small quasiparticle weight $Z_N$
and, therefore, represents a much more sensitive detector of superconductivity.
Strictly speaking, the fact of having a finite $Z_c$ in the thermodynamic
limit does not imply that also $P$ is finite, but only the presence of
paired electrons, as in the example of the two-leg ladder. 
However, in 2D, whenever the compressibility is finite, it is reasonable to
expect condensation of pairs.
Unfortunately, this assumption cannot be supported by direct calculation
in the case of the Hubbard model.
Nevertheless, we can safely discuss the d-wave pairing properties of the
Hubbard model because a non-zero value of $Z_c$ appears very clearly.

Before considering the quantum Monte Carlo results, we show the results
for a small lattice of $L=18$ sites, where it is possible to perform
exact diagonalization.
In Fig.~\ref{small}, we report the results for $Z_c$, $P_d$ and $Z_N$
as a function of $U/t$. The comparison between $P_d$ and $Z_N$
clearly supports our picture: $P_d$ is small in the strong-coupling Hubbard
model because $Z_N$ is small. On the other hand, a much higher signal of paired
electrons can be obtained by studying the pairing strength $Z_c$, which shows
a broad maximum $Z_c \simeq 0.5$ for $U/t\simeq 16$, decreasing to the 
minimum value $Z_c \simeq 0.16$ for $U \to \infty$.
Although the value of the anomalous average $P_d$ and the 
quasiparticle weight are highly overestimated at strong coupling, 
the variational wave function of Eq.~(\ref{wf}) with an optimized $\Delta_k$
provides an excellent estimate of $Z_c$ for all values of $U/t \lesssim 20$, 
thus capturing the correct feature of pairing.

\begin{figure}
\includegraphics[width=0.50\textwidth]{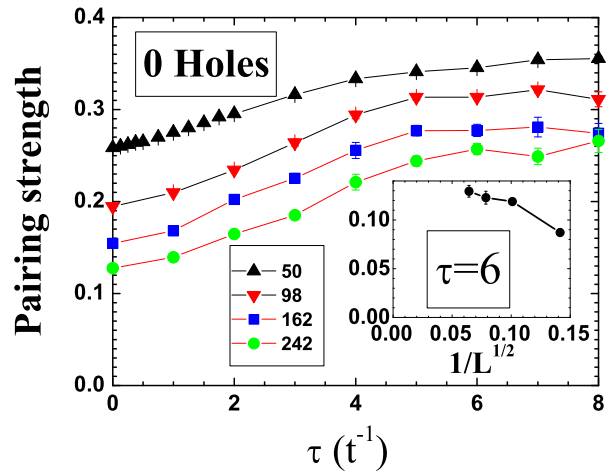}
\vspace{-1.2cm}
\caption{\label{plothalf}
The d-wave pairing strength $Z_c$ as a function of the projection time
$\tau$ for different clusters at half-filling. The inset shows the finite size
scaling of $Z_c-Z_c^G$ for $\tau=6$.
}
\end{figure}

We now turn to larger systems and calculate the pairing strength by using
the zero-temperature Monte Carlo projection technique based on
auxiliary fields.~\cite{sorellahst}
The inclusion of the simple Gutzwiller factor ${\cal J}$, which improves
substantially the convergence in imaginary time $\tau$,
is particularly important because, at finite doping,
the sign problem prevents us to work with arbitrary large imaginary time.

Firstly, we consider the half-filled case. Remarkably, for large systems, 
$Z_c$ increases with the projection time $\tau$, see Fig.~\ref{plothalf}.
In this case, in order to reduce size effects, we have performed the
finite-size scaling of $Z_c-Z_c^G$, where $Z_c^G$ is the value corresponding 
to the Gutzwiller wave function $|\Psi_G \rangle$ with $\Delta \to 0$
($Z_c^G \to 0$ in the thermodynamic limit, see Fig.~\ref{plotbcs}).
Already for $L=50$ sites, the evaluation of the pairing strength is rather
accurate and close to larger sizes.
As shown in Fig.~\ref{plothalf}, $Z_c$ increases monotonically 
with $\tau$, and the value for $\tau=6$ should safely represent, for all 
sizes considered, a rather accurate lower bound for the $\tau\to \infty$ limit.
The thermodynamic limit of $Z_c$ appears therefore clearly finite,
considering also that for $\tau=6$ $Z_c - Z_c^G$ {\it increases}  with
the system size, see the inset of Fig.~\ref{plothalf}.
Our extrapolation is consistent with
a {\it finite} $Z_c \gtrsim 0.1$ in the thermodynamic limit. It is worth
noting that, in this case, a finite value of $Z_c$ does not mean that
the ground state is superconducting, but only that two added
electrons in the half-filled Hubbard model are paired
together in a d-wave singlet. 

\begin{figure}
\includegraphics[width=0.50\textwidth]{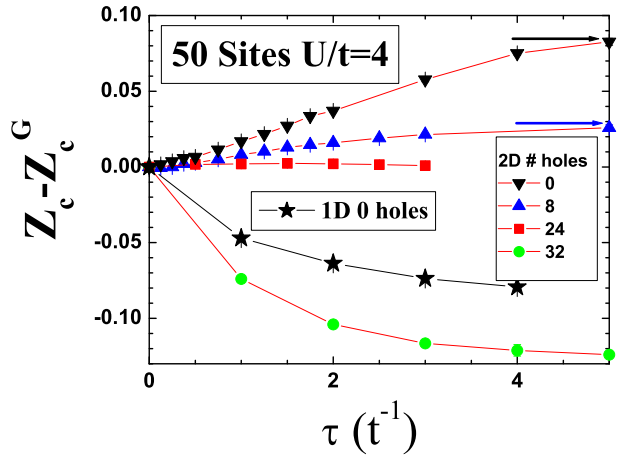}
\vspace{-1.2cm}
\caption{\label{plotdoping}
The d-wave pairing strength $Z_c-Z_c^G$ as a function of the projection time
$\tau$ for different dopings.
The arrows indicate the value of $Z_c$ corresponding to the variational 
wave function of Eq.~(\ref{wf}) with an optimized $\Delta_k$
for $0$ and $8$ holes.
$Z_c$ for the s-wave pairing in one dimension is also shown.
}
\end{figure}

In the doped case the limitation of the sign problem and the strong dependence
upon hole doping prevent us to perform an accurate finite size scaling, 
but, as shown in Fig.~\ref{plotdoping}, the effect remains rather clear, 
and $Z_c-Z_c^G$ increases immediately as soon as the projection time is 
turned on. On the other hand, at large enough doping, we have found
the opposite effect, clearly against d-wave pairing.
Interestingly, also the s-wave pairing strength, obtained by using
an initial wave function $|\Psi_G \rangle$ with s-wave symmetry, 
is not enhanced by the imaginary time projection (not shown).
However, we have not checked other symmetry sectors like p or d$_{xy}$ 
and, therefore, at large doping there maybe other types of pairing 
instabilities.~\cite{hlubina}
It is remarkable that also for $50$ sites the simple Gutzwiller wave function 
with a finite d-wave gap function (stabilized in the underdoped region)
describes accurately the enhanced pairing strength $Z_c$, as shown by 
the arrows in Fig.~\ref{plotdoping}. 

Although we worked at zero temperature, one should be tempted to associate
$Z_c$ with the pseudogap observed in the HTc superconductors. In this 
scenario, the onset of the pseudogap, for a temperature $T^*$, is marked 
by a finite value of $Z_c$, indicating paired electrons without phase 
coherence, then at a much lower temperature $T_c$ the pairs eventually 
condense, giving rise to a true long-range order.~\cite{patricklee}

In conclusion, it turns out that the pairing is a robust property of 
lightly-doped Mott insulators and appears already in small size calculations.
The pairing strength $Z_c$ increases with decreasing doping and has its
maximum at half filling, where phase coherence is inhibited by the charge gap.
Notice that this is a peculiar feature of the 2D model, as in one dimension 
no evidence of a finite pairing strength is found even at half filling, see 
Fig.~\ref{plotdoping}.
We argue that a finite pairing strength for an insulator is 
just the qualitative new feature that discriminates a band insulator 
from a 2D RVB insulator, which is defined in terms of singlet pairs,
as in a fully projected BCS wavefunction of Eq.~(\ref{wf}).
Within this definition, determined by the measurable quantity $Z_c$,
and {\it independent of the variational ansatz}, we discover the possibility 
to have an RVB-like insulator also when the existence of a finite pairing 
strength is accompanied by the antiferromagnetic long-range order, like in 
the half-filled Hubbard model.~\cite{scalettar} 
In this simple model, the insulator is somehow prepared 
to become superconductor (with d-wave symmetry) and 
a small amount of doping allows the propagation of the RVB pairs.~\cite{rvb}
This scenario offers a simple and natural explanation 
of the ultimate mechanism of HTc superconductivity.

We acknowledge useful discussion with R. Hlubina, M. Fabrizio, and A. Parola. 
In particular we thank Shiwei Zhang for fruitful discussions and for sharing 
with us unpublished results. This work was partially supported by 
MIUR (COFIN 2003). F.B. is supported by INFM.

\end{document}